\documentclass[onecolumn]{aastex63}

\usepackage[utf8]{inputenc}
\usepackage[a4paper, total={6in, 8in}]{geometry}
\usepackage{natbib}

\usepackage{amsmath}
\usepackage{mathtools}
\usepackage[toc,page]{appendix}

\usepackage{graphicx}
\usepackage{esvect}

\def\deg{$^\circ$}

\newcolumntype{L}{>{\centering\arraybackslash}m{2.5cm}}

\shorttitle{An Analytic Model for Exoplanets}
\shortauthors{Judkovsky et al.}

\begin{document}
\received{August 30, 2021}

\title{An Accurate 3D Analytic Model for Exoplanetary Photometry, Radial Velocity and Astrometry}

\correspondingauthor{Yair Judkovsky}
\author[0000-0003-2295-8183]{Yair Judkovsky}
\affiliation{Weizmann Institute of Science, Rehovot 76100 Israel}
\email{yair.judkovsky@weizmann.ac.il}

\author[0000-0002-9152-5042]{Aviv Ofir}
\affiliation{Weizmann Institute of Science, Rehovot 76100 Israel}

\author[0000-0001-9930-2495]{Oded Aharonson}
\affiliation{Weizmann Institute of Science, Rehovot 76100 Israel}
\affiliation{Planetary Science Institute, Tucson, AZ, 85719-2395 USA }

\graphicspath{{./}{Figures/}}


\begin{abstract}
We developed and provide \texttt{AnalyticLC}, a novel analytic method and code implementation for dynamical modeling of planetary systems, including non-coplanar interactions, based on a disturbing function expansion to 4th order in eccentricities and inclinations. \texttt{AnalyticLC} calculates the system dynamics in 3D and the resulting model light curve, radial velocity and astrometry signatures, enabling simultaneous fitting of these data. We show that for a near-resonant chain of three planets, where the two super-periods are close to each other, the TTVs of the pair-wise interactions cannot be directly summed to give the full system TTVs because the super-periods themselves resonate. We derive the simultaneous 3-planets correction and include it in \texttt{AnalyticLC}.
We compare the model computed by \texttt{AnalyticLC} to synthetic data generated by an N-body integrator, and evaluate its accuracy. Depending on the maximal order of expansion terms kept, \texttt{AnalyticLC} computation time can be up to an order of magnitude faster than state-of-the-art published N-body integrator \texttt{TTVFast}, with smaller enhancement seen at higher order. The advantage increases for long-term observations as our approach's computation time does not depend on the time-span of the data. Depending on the system parameters, the photometric accuracy is typically a few ppm, significantly smaller than typical {Kepler's} and other observatories' data uncertainty. 
Our highly efficient and accurate implementation allows full inversion of a large number of observed systems for planetary physical and orbital parameters, presented in a companion paper.
\end{abstract}


\section{Introduction}
Since the first discoveries of extrasolar planets \citep{Latham89, Wolszczan1992, Mayor1995}, our knowledge of planetary systems is continuously growing. To date, more than 4,400 confirmed extra-solar planets are known. The accumulating knowledge of the physical and orbital characteristics of individual planets and systems helps us to better understand the collective nature of planetary systems - which were found to be a natural and common consequence of star formation .

While the first detections of exoplanets were done mainly using the Radial Velocity (RV) method, the {Kepler} mission \citep{BoruckiEtAl2010} provided a great leap in the number of planets detections by utilizing the transit method, yielding about 2,300 new confirmed planets. Due to the large potential of these data sets, which continue to rapidly grow nowadays with the Transiting Exoplanet Survey Satellite (TESS) mission \citep{RickerEtAl2010} and in the future PLAnetary Transits and Oscillations of stars (PLATO) mission \citep{RauerEtAl2014}, efforts have been made in order to develop efficient methods interpretation of these data sets.

The most direct way to invert RV or light-curve data to physical parameters is by running multiple N-body integrations driven by some non-linear fitting process. Such a method was used for RV data \citep[e.g.][]{RiveraEtAl2005} and for photometric data  \citep[e.g.][]{MillsFabrycky2017, MillsEtAl2019, FreudenthalEtAl2018, GrimmEtAl2018} to fit for the parameters of a several planetary systems. A drawback of this method is that as the observational time span increases, the modeling time becomes longer, making this a cumbersome, and sometimes even prohibitive, process. 

In order to make the process of interpreting long-baseline photometric data more efficient, analytic methods were developed for data interpretation. In addition to the efficiency of these methods, they are powerful in providing us with better understanding of the physical processes that govern the dynamics of planetary systems.

Since the time-of-mid-transit is usually a well-constrained property of an individual sufficiently deep transit event, TTVs (Transit Timing Variations, \citet{AgolSteffenSariClarkson2005}), were a focus of many studies that use methods to invert them for planetary parameters - and planetary mass in particular. \citet{NesvornyMorbidelli2008} used perturbation theory to analytically derive the TTV. This method was extended to eccentric and inclined orbits \citep{Nesvorny2009} and then implemented and used in the code TTVIM \citep{NesvornyBeauge2010}. \citet{Meschiari2009} developed an TTV-RV joint fitting software and used it to study the potential of their joint analysis \citep{Meschiari2010}, and \citet{Payne2010} studied the magnitude of the TTV effect for various inclination regimes. 
\citet{LithwickXieWu2012} explained the observed long-period TTV ("super-period") near first order MMR (Mean Motion Resonance) as the period of circulation of the line of conjunction between two interacting planets. The interpretation of the system parameters from this TTV pattern is impacted by a mass-eccentricity degeneracy, which can sometimes be broken by detecting the high-frequency TTV components, also called "chopping TTV" \citep{NesvornyVokrouhlicky2014,DeckAgol2015}, or by detecting TTV to second order in eccentricity \citep{HaddenLithwick2016}.
 
The ability of the TTV modeling to invert for the planetary masses led to the development of open-source software tools dedicated to model TTVs either by a full N-body integration \citep{DeckAgolHolmanNesvorny2014} or by analytic calculation to first order in eccentricities \citep{AgolDeck2016}. \citet{LinialGilbaumSari2018} constructed a modal decomposition method using a geometric approach to invert TTV data -- also accurate to first order in eccentricity. Because of the abundance of TTVs in the Kepler population \citep{HolczerEtAl2016, OfirEtAl2018}, TTV data were used to extract masses and eccentricities (or combinations of them) for a large number of planets, e.g. \citet{HaddenLithwick2016, HaddenLithwick2017, JontofHutter2021, Yoffe2021}, and many more.

While in many cases fitting times-of-mid-transit can yield good understanding of planetary properties, this method does have drawbacks. As highlighted by \citet{OfirEtAl2018}, fitting transit times rather than flux does not fully exploit the information encoded in the light-curve, because many degrees of freedom are required in order to translate the full light-curve to individual transit times. Fitting transit times also creates a bias that favors large planets (of clear transits that enable good estimate of the mid-transit time) with strong TTV signals - while often the desired signals are those of small planets with shallow transits.

A second drawback of fitting times rather than flux is that it reduces the light-curve data to mid-transit-times alone, erasing valuable information encoded in other types of transit variations, such as depth and duration. These are of great interest, as they can probe non-coplanar interactions within the system. 

Both the transit and RV methods are limited in the ability to assess mutual 3D inclination: RV is inherently blind to inclination, while transit is inherently biased to selecting flat multi-transiting systems \citep{RagozzineHolman2010}, and is symmetric to rotation about the line of sight. Therefore, many studies used a statistical approach to assessing the mutual inclinations. Such an approach aims to characterize the {\it dispersion} of mutual inclinations (and/or eccentricities) rather than infer from the values of individual systems \citep{FabryckyEtAl2014}. \citet{XieEtAl2016} have shown that the excess of singly transiting planets can be explained by the a bimodal distribution which is characterized by different dispersion of mutual inclinations and eccentricities for "hot" and "cold" systems. The dichotomic explanation for the dispersion of eccentricities and inclinations was further studied by \citet{He2019}. These authors developed a framework for simulating planetary populations, and used it to fit a bi-modal distribution to the Kepler planet population, taking into account a large set of observable quantities. In a following work \citep{He2020}, this dichotomic explanation for the excess of singly-transiting planets has been revised by suggesting a non-dichotomous model based on Angular Momentum Deficit (AMD, \citet{Laskar2017}). The AMD-stable model provides a more natural and physically-based theory than the dichotomic model. \citet{Millholland2021} performed a statistical analysis of the number of observed transit duration variations (TDVs) \citep{Shahaf2021} to show that the AMD-based model is statistically preferred over a dichotomous model. Another point of view of the mutual inclinations within planetary systems was addressed be \citet{Masuda2020}, who studied the mutual inclinations between inner rocky planets and outer giants by the occurrence rate of transits of cold Jupiters and their general abundance based on RV data. They find that systems with multiply-transiting inner rocky planets are expected to posses lower mutual inclination with their outer giants than singly-transiting inner rocky planets. 

These examples show how statistical methods can infer on the general dispersion of mutual inclinations; however, constraining the mutual inclination of specific systems has been done only for a handful of cases \citep[e.g.][]{MillsFabrycky2017}. It is of high interest to pin-point systems which are likely to posses mutual inclinations, as this will enable focusing the observational efforts. A sign of mutual inclinations is the existence of TDVs, or impact parameter variations (TbVs), since for low-eccentricity orbits $b$ and $D$ are more sensitive to nodal precession than to apsidal precssion. In order to detect such signals in the data, which are typically weak and slow, a global light-curve model that contains both TTVs and TbVs should be utilized. A model that jointly fits TTVs and TbVs is a key to probing the existence of mutual inclinations in individual planetary systems.

Here we present \texttt{AnalyticLC}, an analytic tool that accurately and efficiently  calculates a full light-curve model, based on planetary physical and orbital parameters. The basic characteristics of the tool that guided its development are: (i) Full light-curve modeling, exploiting the photometric data (as opposed to just transit times). (ii) Computational speed to enable multi-dimensional inversions. (iii) Analytical treatment that sheds light on the dynamical processes manifested in the data. (iv) Applicability to multiple data types, including photometry, RV and astrometry.

This paper is organized as follows: In \S\ref{s-TheModel} we describe the mathematical model of a single transit event, and link it to the instantaneous orbital elements. In addition, we present analytic formulae for TTVs that arise from the simultaneous interactions of three planets on a near-resonant chain with each other, a frequent scenario in the {Kepler} population. In \S\ref{s-TestingTheModel} we test the accuracy limits of the model against full N-body integrations for 2- and 3- planet systems, and conclude in \S~\ref{sec:Summary}. In appendix \S\ref{app-A} we give the full mathematical derivation of the orbital elements, and in appendix \S\ref{app-B} we describe some code technicalities that enhance the calculation speed.

\section{The model}\label{s-TheModel}

\subsection{Overview}

The model is based on constructing an approximate solution of Lagrange's planetary equations of motion, and then translating the orbital elements values to transit properties. Once the properties of each individual transit are known, the instantaneous flux is calculated using the Mandel-Agol formula \citep{MandelAgol2002}. 
First, the disturbing function is expanded to the desired order in $e, I$ (up to fourth order in \texttt{AnalyticLC}), and Lagrange's planetary equations of motion are written \citep{SSD1999}. The secular terms ({\it i.e.}, terms that do not depend on mean longitudes) are isolated, and the equations are solved for them analytically by matrix inversion \citep[chapter 7]{SSD1999}. This yields the so-called free eccentricity \citep{LithwickXieWu2012}, and the corresponding ``free inclination". We note that the division to ``free" and ``forced" as used here follows the definition of \citet{LithwickXieWu2012}, not to be confused with that used by \citet{SSD1999} to describe two time scales in the motion of a test particle experiencing secular perturbations. Our treatment of the free elements differs from the one presented by \citet{LithwickXieWu2012} in the sense that we allow secular motion, and hence the free values change in time, while in their method the free values are assumed constant in time. After solving for the secular motion of the free values, the equations are written for the non-secular terms only (these include both synodic terms and resonant terms), and they are solved by assuming that their right hand sides are time-dependent only through the mean longitudes, which vary on an orbital time scale. The other elements are assumed to be equal to their free values for the sake of integrating these equations. This technique is similar to the one applied by \citet{HaddenLithwick2016}, who used the variations in the orbital elements to construct analytic expressions for TTV to second order in eccentricity under the assumption of co-planarity. Similar to their approach, here we expand on the TTV analysis by calculating the variations in all orbital elements and translating them to transit properties of each individual event, in addition to incorporating secular effects which give rise to slow, gradual variations in the transit shape. 

Throughout the derivation we use astrocentric coordinates, as these are natural for describing a transit event; only for the computation of RV/Astrometry model we transform to barycentric coordinates. In appendix \S\ref{app-A} we give the detailed analytic derivation. The general steps of the \texttt{AnalyticLC} model are summarized in table-\ref{table:MethodSteps}.

\begin{table}
{\centering
\begin{tabular}{|l|p{3cm}|p{3cm}|p{4cm}|p{3cm}|} \hline
\#& Step & Technique & Assumptions & Outputs   \\ \hline\hline
 1 & Secular orbital motion   &      matrix inversion & maintaining secular terms up to $\mathrm{2^{nd}}$ order in $e,I$ &               motion of free eccentricities and inclinations  \\ \hline
 2 & Near-MMR orbital motion   & analytic approximate integration of Lagrange's equations of motion & maintaining resonant terms up to $\mathrm{4^{th}}$ order in $e,I$; slow motion of all elements except mean longitudes & near-resonant motion of all elements                   \\ \hline 
 3 &  Individual transit properties (\S\ref{sec:GeometryToProperties})  & geometric calculation & along the transit, the local shape of the orbit is approximated by a circular arc & individual transit parameters                   \\ \hline 
 4 &  Light-curve generation   & Mandel-Agol model;  binning & Mandel-Agol model with two limb-darkening parameters & full light-curve            \\ \hline
\end{tabular}
\caption{The general flow of \texttt{AnalyticLC}. The first two steps relate to the system dynamics; the third step translates the orbital motion to transit parameters; the fourth step computes a full light-curve. RV and astrometry values are also calculated for any required time.}
\label{table:MethodSteps}
}
\end{table}

\subsection{Orbital Geometry to Transit Properties}
\label{sec:GeometryToProperties}

Let us begin by expressing the transit properties using the orbital elements. For a circular orbit at an orbital period $P$, with an orbital radius $a$, a normalized impact parameter $b$ and mean motion $n=2\pi/P$, the sky-plane coordinates around the transit are given by

\begin{equation}
    y = \frac{a}{R_*}\sin\phi
\end{equation}
and

\begin{equation}
     z = b\cos\phi,
\end{equation}
where the origin is at the center of the star, $x$ points at the observer, $y,z$ form the plane of the sky, the phase is given by $\phi=n(t-t_{\rm mid})$, $R_*$ is the stellar radius, $t$ is the time and $t_{\rm mid}$ the time of mid-transit. 

Inverting the relation between sky-position and time yields the transit duration $T$ (time between the two points at which the distance between the centers of planet and star is 1 stellar radius) and the ingress/egress time $\tau$, as following:

\begin{equation}
    T = \frac{2}{n}\arcsin{\sqrt{\frac{1-b^2}{a^2-b^2}}}
\end{equation}
and
\begin{equation}
    \tau = \frac{1}{n}\left|\arcsin\sqrt{\frac{(1+r)^2-b^2}{a^2-b^2}}-\arcsin\sqrt{\frac{(1-r)^2-b^2}{a^2-b^2}}\right|.
\end{equation}

Similar analytic expressions for the circular orbit case were given by \cite{SeagerMallen-Ornelas2003}.

For an elliptic orbit the value of $a$ is replaced by the planet-star separation at mid-transit, and $n$ is replaced by the angular velocity at mid-transit, respectively:
\begin{equation}
    d=\frac{a(1-e^2)}{1+e\cos{f}}
\end{equation}
and
\begin{equation}
    \dot{\theta}=n\frac{(1+e\cos{f})^2}{(1-e^2)^{3/2}},
\end{equation}
where $e$ is the eccentricity and $f$ is the true anomaly at mid-transit. The impact parameter for an elliptic orbit is given by

\begin{equation}
    b = \frac{a/R_*(1-e^2)}{1+e\cos{f}}\cos{i},
\end{equation}
where $i$ is the inclination with respect to the plane of the sky.

Assuming that along the transit one can treat the trajectory as a small arc of a circular path with a constant angular velocity (justified provided that $T\ll{}P$), one can calculate the stellar-planet sky-projected separation at any given time around mid-transit. The light-curve is obtained from this distance using the Mandel-Agol formula \citep{MandelAgol2002}.

In order to conveniently calculate the transit timing variations, we work in an axis system where the $x$-axis points at the observer, similar to previous works, e.g. \citet{LithwickXieWu2012, HaddenLithwick2016}. In this case, the mid-transit true longitude is 0, which allows for calculating the mid-transit times conveniently. The $y$ axis is defined such that the longitude of ascending node, $\Omega$, of the innermost planet, lies on the positive direction of $y$. The sky is the plane $yz$. The orbital inclination $I$ is the angle between the orbit normal and the $z$ axis. 
Under these conventions, the true anomaly at mid-transit is  $f=-\varpi$, and hence $\cos{f}$ in the expressions above is replaced by $\cos\varpi$. The inclination of the orbit normal with respect to the line of sight $i$ is a projection of the angle $I$ between the $z$ axis and the orbit normal, specified in azimuth by $\Omega$ (the longitude of ascending node). Thus, 
\begin{equation}
    \cos i = \sin I\sin\Omega.
\end{equation}
For a transit to occur, $i$ should be close to 90 degrees.

\begin{figure}[h]
    {\includegraphics[width=.9\linewidth]{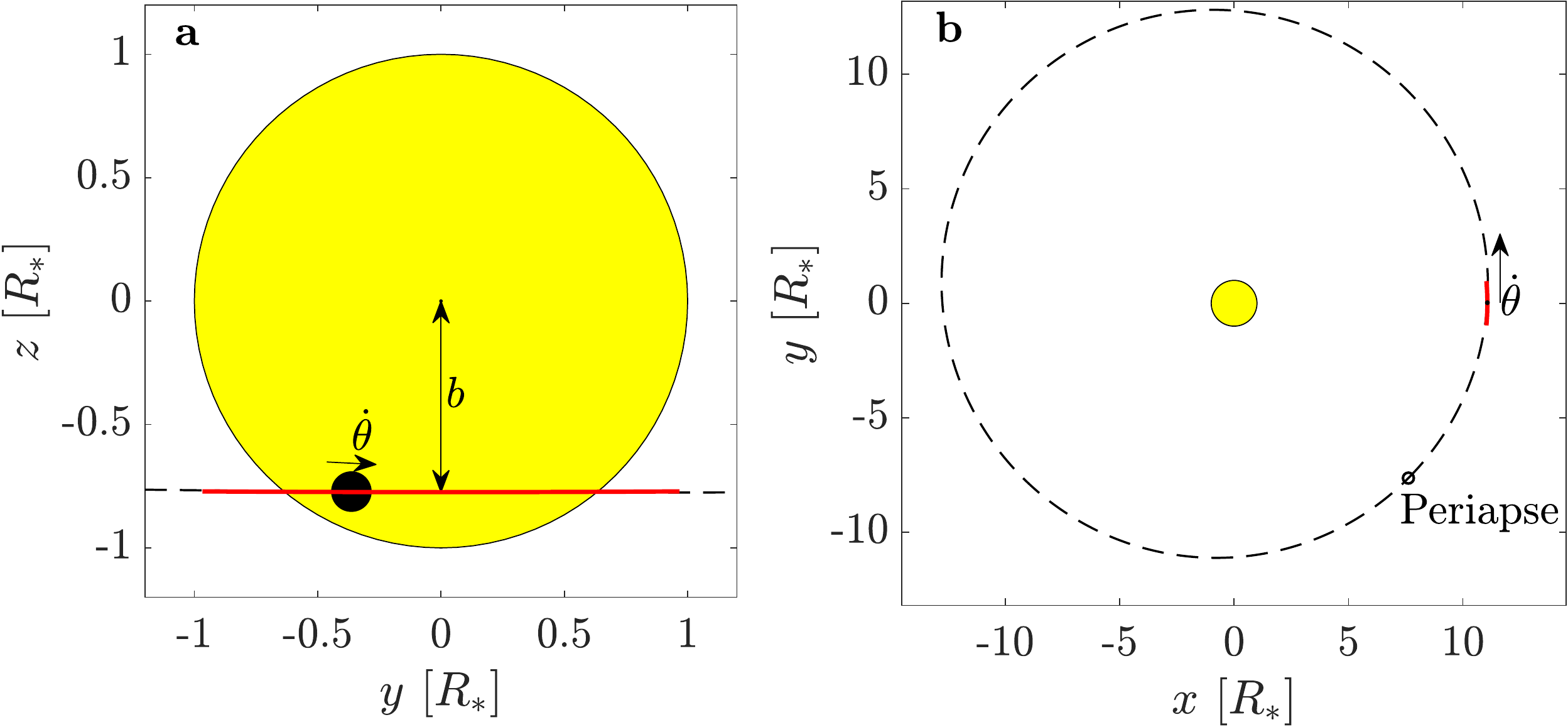}}
    {\caption{The orbital geometry and orbital parameters of a transit event, here illustrated for $e=0.1$. In both panels the dashed line is the orbit, and the red solid lines are the circular arc approximation. (a) The orbit projected on the sky-plane, with impact parameter $b$ and planetary angular velocity $\dot{\theta}$. (b) The $xy$ plane. $d/R_*$ is the star-planet separation at mid-transit, in stellar radii - due to the orbital inclination in this case, the line connecting the planet and star is not in the $xy$ plane. The $x$ axis points at the observer, $y$ is along the direction of motion at mid-transit.}
    \label{fig:OrbitalGeometryIllustration}}
\end{figure}

The transformations above enable translating the instantaneous orbital elements at mid-transit to a full description of the transit shape as a function of time. The orbital geometry is illustrated in Fig.~\ref{fig:OrbitalGeometryIllustration}.

\subsection{Orbital Elements Variations}
The formulae above show how a set of the osculating elements $a,e,\varpi,I,\Omega$ can fully describe the shape of any single transit under the assumption that they all vary slowly enough, relative to the transit duration, such that they are locally constant. The 6th orbital element, the mean longitude $\lambda=M+\varpi$ (where $M$ is the mean anomaly), affects the transit time, but not its shape. The problem of generating a full light-curve model is then reduced to calculating the orbital elements as a function of time, on time scales longer than $T$. We approximate them by using the disturbing function formalism. This has been done for the calculation of TTVs via the small variations in $\lambda$ and $z=e\, \exp{(i\varpi)}$ to second order in eccentricities, by \cite{HaddenLithwick2016}, under the assumption of coplanarity. We extend this method to calculating the variations of all orbital elements to fourth power in eccentricities and inclinations and in 3D. We summarize the method here; the full derivation is given in Appendix \S\ref{app-A}.

The potential energy of the gravitational interactions among the planets can be described as a series sum over different frequencies. The secular terms, which do not depend on the mean longitudes, cause a drift in the eccentricity and inclination vectors - a slow motion with a frequency proportional to the orbital frequency of the perturbed planet and to the mass ratio of the perturber and the host star. Superimposed on this slow secular drift, the eccentricity and inclination vectors oscillate with various frequencies, including  the strongest component at the inverse super-period, arising from the nearest first-order-MMR \citep{LithwickXieWu2012}. The full motion can be divided to near-resonant frequencies (as will be explained below, we do not treat the case of resonance locking) and synodic frequencies, that cause the so-called "chopping" effect \citep{DeckAgol2015}.
The flow of \texttt{AnalyticLC} is composed of four steps, and incorporates both secular and near-MMR dynamics, described in Table~\ref{table:MethodSteps}.

An underlying important assumption in the approximate solution of Lagrange's equations is that the conjunction longitude circulates rather than librates; in other words, the system is not locked in resonance. The out-of-resonance configuration is the overwhelmingly common case for the observed {Kepler} population. This observational finding has been explained by dissipation processes, which tend to push resonance-locked systems wide of resonance  \citep[e.g.][]{LithwickWu2012a, Millholland2019}.

\subsection{Three-planets interactions}\label{sec:3Planets}

Past works that analytically calculated the TTVs treated them as perturbations to first order in the perturber mass. In such a description, the perturbations are additive, and hence the total TTV that a planet experiences is the sum of TTVs applied by the different individual perturbers.
In Appendix \ref{app-A} we show that for a triplet of planets that constructs a chain of two near-first-order-MMR's, new TTV patterns can arise, not negligible in comparison to the first-order effect. In fact, if the two super-periods of the near-resonant-chain are similar, then this additional TTV amplitude can reach the amplitude of the individual near-1st-order MMR TTVs, which were introduced by \cite{LithwickXieWu2012}. The additional TTV pattern is of second order in the planetary masses, but it also scales inversely with the sum or difference between the super mean motions. In other words, if the super-periods match, they resonate themselves. We refer to this phenomenon as Super-Mean-Motion Resonance (SMMR), because it involves a resonance between the super mean motions. 
Explicitly, if the inner planets are near a $j:j-1$ resonance, and the outer pair is near a $k:k-1$ resonance, then the additional TTV is given by these expressions:
\begin{eqnarray}
    \delta t^{(2-3)} = & \frac{P}{2\pi}\frac{3}{2}\frac{m'm''}{(m_*+m)(m_*+m')}n'n^2\alpha_{12}\alpha_{23}\frac{j-1}{n^k}f_{27}(\alpha_{23})(f_{31}(\alpha_{12})-2\alpha_{12}\delta_{j,2}) \nonumber \\
    & \left(\frac{\sin{(\lambda^j+\lambda^k-2\varpi')}}{(n^j+n^k)^2}-\frac{\sin{(\lambda^k-\lambda^j)}}{(n^j-n^k)^2}\right),
\end{eqnarray}
\begin{eqnarray}
    \delta t'^{(1-3)} = & \frac{P'}{2\pi}\frac{3}{2}\frac{mm''}{(m_*+m')^2}n'^3\alpha_{23}f_{27}(\alpha_{23})(f_{31}(\alpha_{12})-\frac{\delta_{j,2}}{2\alpha_{12}^2}) 
    \nonumber \\
    & \left( (\frac{j}{n^k}+\frac{1-k}{n^j})\frac{\sin{(\lambda^j+\lambda^k-2\varpi')}}{(n^j+n^k)^2}+(\frac{j}{n^k}-\frac{1-k}{n^j})\frac{\sin{(\lambda^j-\lambda^k)}}{(n^j-n^k)^2}\right),
\end{eqnarray}
and
\begin{eqnarray}
    \delta t''^{(1-2)} = & \frac{P''}{2\pi}\frac{3}{2}\frac{mm'}{(m_*+m)(m_*+m'')}n'n''^2\frac{k}{n^j}f_{27}(\alpha_{23})\left(f_{31}(\alpha_{12})-\frac{\delta_{j,2}}{2\alpha_{12}^2}\right) \nonumber \\
   & \left(\frac{\sin{(\lambda^j+\lambda^k-2\varpi')}}{(n^j+n^k)^2}-\frac{\sin{(\lambda^j-\lambda^k)}}{(n^j-n^k)^2}\right),
\end{eqnarray}
where $m$ is the planetary mass, $m_*$ is the stellar mass, $P$ is the orbital period, $n$ is the mean motion, $\alpha_{ij}$ is the ratio between the semi-major axes (of planets $i$ and $j$), $f_{27}, f_{31}$ are functions of the Laplace coefficients, given by \citet[Appendix 2B]{SSD1999}. The unprimed quantities refer to the innermost planet (1), the primed quantities refer to the intermediate planet (2), and the double-primed quantities refer to the outermost planet (3). The upper scripts represent the planets that generate the cross interaction and on which the additional TTV depends; for example, $\delta t^{(2-3)}$ is proportional to the masses of planets 2 and 3. The frequencies $n^j$ and $n^k$ are the super-mean-motions, related to the super-periods of the near-resonant interactions \citep{LithwickXieWu2012}. They are given as a function of the orbital mean motions by
\begin{equation}
    n^j=jn'+(1-j)n
\end{equation}
and
\begin{equation}
    n^k=kn''+(1-k)n'.
\end{equation}
The full derivation for this TTV pattern is given in Appendix \ref{app-A}.
 
The frequency of the 3-planet-interaction TTV is equal to the sum, or difference, between the super mean motions of the two pairs. If the super periods are too close and this frequency is too small relative to the duration of the observation, this TTV pattern will not be seen. In other words, if the observation period is less than $\approx$0.25 of the 3-planets-TTV period, the curvature of the sine would not be caught. For a given inner super-period $P^j$ and an observation duration $D$, The non-detection limits for the 3-planet TTV as a function of the outer pair super-period $P^k$ are given by the conditions
\begin{equation}
\label{eq:DetectionLimit}
    \frac{4DP^j}{4D+P^j} \lessapprox P^k \lessapprox\frac{4DP^j}{4D-P^j}
\end{equation}
where the factor 4 in both of these equations come from the requirement that the observation time would be 4 times larger than the period of the 3-planets effect; this is a heuristic requirement as the exact factor would depend on the exact phase in which the observation is performed.
 
\begin{figure}
    {\includegraphics[width=.9\linewidth]{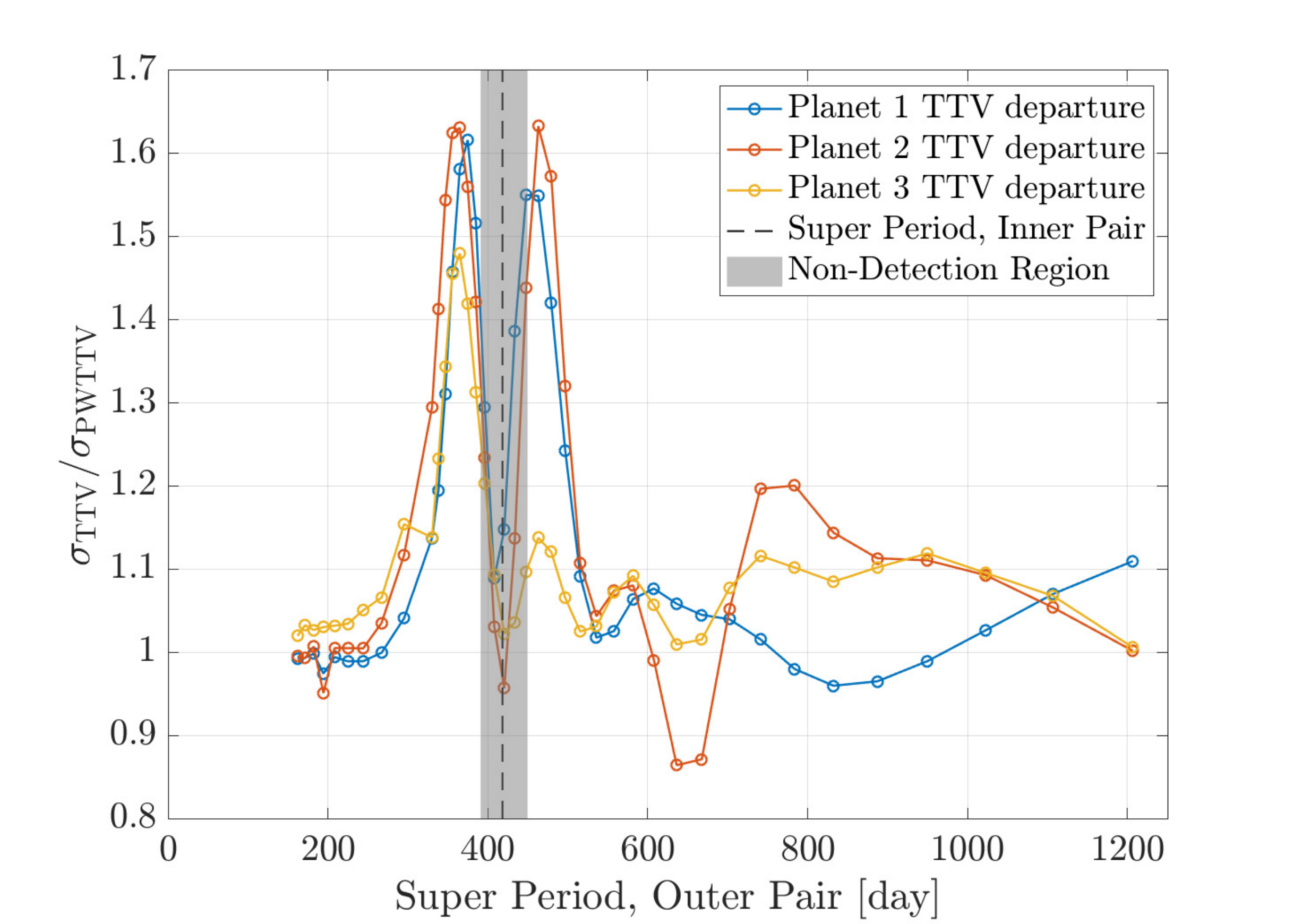}}
    {\caption{Relative departure of the 3-planets TTV magnitude ($\sigma_{\rm TTV}$) from the pairwise TTV magnitude ($\sigma_{\rm PWTTV}$). This figure shows that for a near-SMMR (\S\ref{sec:3Planets}) configuration, the pair-wise TTV approximation breaks, and that Equation~\ref{eq:DetectionLimit} gives a good prediction on the regions at which the additional TTV pattern should be detectable. Blue, orange and red circles represent the relative deviation of the innermost, intermediate and outer planets TTV magnitudes from the pair-wise calculation as a function of the outer pair super-period. The inner pair super-period is indicated by a black dashed line. Within the gray region this deviation is not observable in the given time span (Equation~\ref{eq:DetectionLimit}). }
    \label{fig:TTV3DetectionLimits}}
\end{figure}

Between these two limit, the phenomenon would be small. This is illustrated in Figure~\ref{fig:TTV3DetectionLimits}. In this figure we show the results of a computational experiment involving a system of three planets in a near-MMR chain.
The orbital periods of the inner pair are kept constant such that their super-period is $\approx$418 days, while the orbital period of the outer planet varies such that the super-period of the outer pair varies between 160 and 1200 days. For each orbital configuration, the total TTVs are calculated by a single 3-planets N-body integration, and for each planet the standard deviation of the arising TTV, $\sigma_{\rm TTV}$, is calculated. The pair-wise TTVs are calculated using three 2-planets integrations, and the standard deviation of this pair-wise calculated TTV is denoted $\sigma_{\rm PWTTV}$. The quantities $\sigma_{\rm TTV}$ and $\sigma_{\rm PWTTV}$ reflect, in one number per orbital configuration, the magnitude of the TTV for the simultaneous 3-planets interaction and for the pair-wise interaction. Throughout all integrations, we kept the free eccentricities and inclinations constant by performing a fine adjustment to the initial conditions. When the super-period of the outer pair approaches that of the inner, the TTVs of all three planets deviate from the pair-wise calculation. When the super-periods are close enough to each other such that the SMMR effect is of too low a frequency for the observation time to capture, the effect is not seen in the TTV (it is absorbed by an adjusted mean period). The theoretical region at which the SMMR-TTV should not be detected (Equation~\ref{eq:DetectionLimit}), matches nicely the results of the numerical experiment.
 
\section{testing the model}\label{s-TestingTheModel}

In this section, we present a few of the tests we performed on the model to check its validity and its accuracy limits. 

\subsection{Comparison with N-body - 2 Planets}
The first test presented here shows the comparison of the analytic model with an N-body integration of a 2-planets system near the first-order 3:2 MMR with small eccentricities inclinations. The N-body integration was done using \texttt{MERCURY6} \citep{Chambers1999}, which gives as an output the full set of orbital elements as a function of time. The mid-transit times were estimated by using a weighed Keplerian arcs for the two points bracketing the transit, as done in the code \texttt{TTVFast} \citep{DeckAgolHolmanNesvorny2014}.
In table \ref{table:SimulatedSystem}, the main parameters of the modeled system are presented. The integration time span is 1500 days with a rather oversampled time steps of 0.5 hours, similar to {Kepler}'s long cadence \citep{BoruckiEtAl2010}.

\begin{table}
{\centering
\begin{tabular}{|l||l|l|l|l|l|l|} \hline
Parameter  & $P$ [day] & $a$ [AU] &     $e$ & $I$ [\deg] & $m~[m_\oplus]$  &  $R~[R_\oplus]$\\ \hline\hline
Planet 1   &  11.551   &      0.1 & 0.014 & 1.41 &               6 &         2.4495 \\ \hline
Planet 2   &  17.683   &   0.1328 & 0.014 & 3.04 &               9 &             3  \\ \hline 
\end{tabular}
\caption{The main parameters of the simulated 2-planets system. The two planets are near the 3:2 MMR, with a period ratio of about 1.531, equivalent to a normalized distance from resonance of $\approx 0.02$.}
\label{table:SimulatedSystem}
}
\end{table}

\begin{figure}
{\includegraphics[width=1\linewidth]{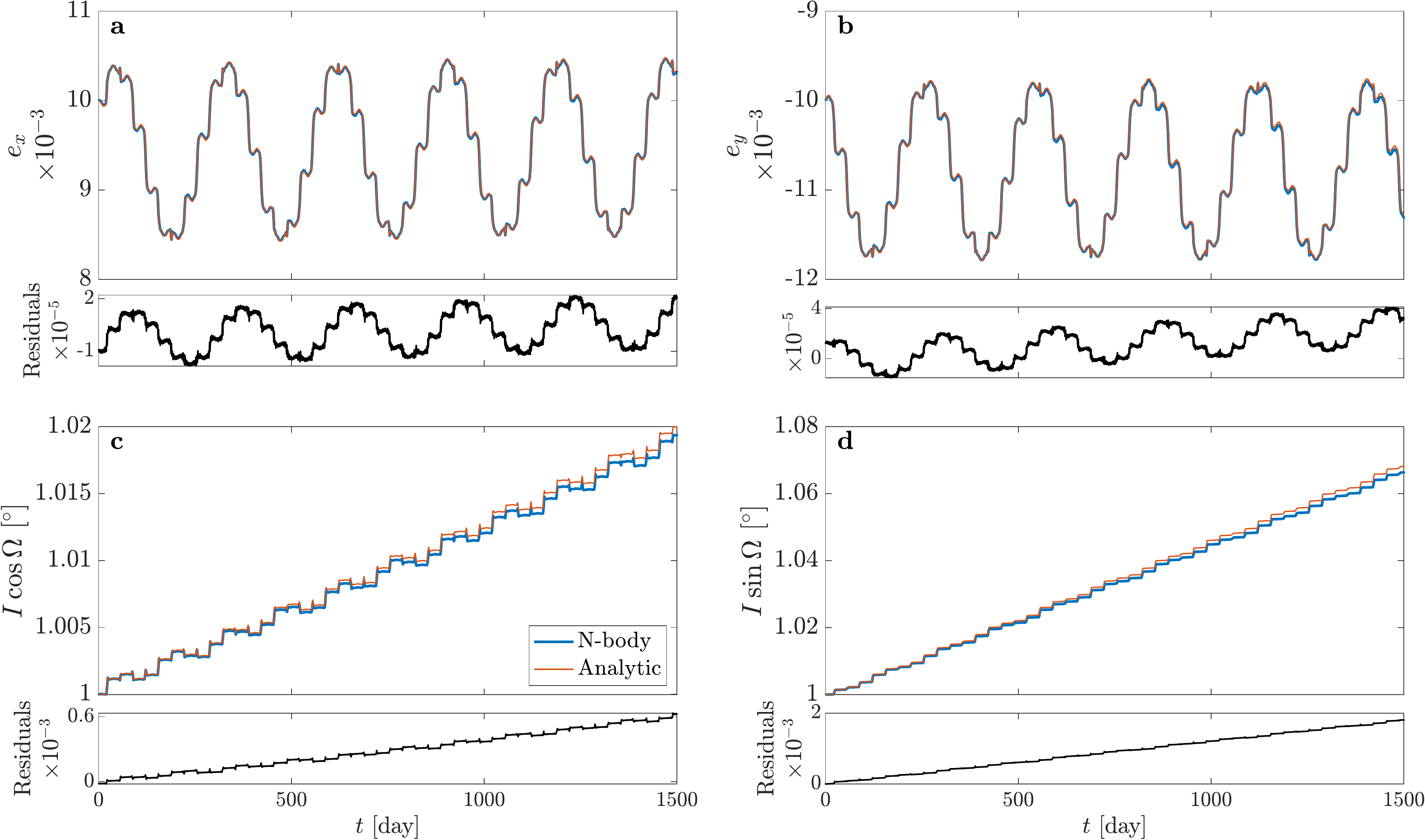}}
    \caption{Motion of the inner planet orbital elements along the simulation time span, demonstrating the ability of \texttt{AnalyticLC} to capture both long-term and short-term variations in the orbital elements. Blue curves show the result of an N-body integration, orange show the result obtained from \texttt{AnalyticLC}. Residuals are shown in black. (a,b) Eccentricity vector components motion. (c,d) inclination vector components motion. }
    \label{fig:OrbitalElementsMotion}
    
\end{figure}

In Figure~\ref{fig:OrbitalElementsMotion} we show the accuracy \texttt{AnalyticLC} for the eccentricity and inclination calculation vectors motion for the inner planet by comparing them to their values from an N-body integration performed by MERCURY6 \citep{Chambers1999}. In order to perform the comparison, we translated the osculating orbital elements at which the N-body integration was initiated to the free elements which are used by \texttt{AnalyticLC} by calculating the forced elements and subtracting them from the osculating ones, as described in \ref{app-A}.  The eccentricity vector variations are dominated by the forced motion due to the near-resonant interaction, while the slow variation of the inclination is a result of secular interactions. The residuals are about two orders of magnitude smaller than the amplitudes of the orbital element variations.

\begin{figure}
{\includegraphics[width=.9\linewidth]{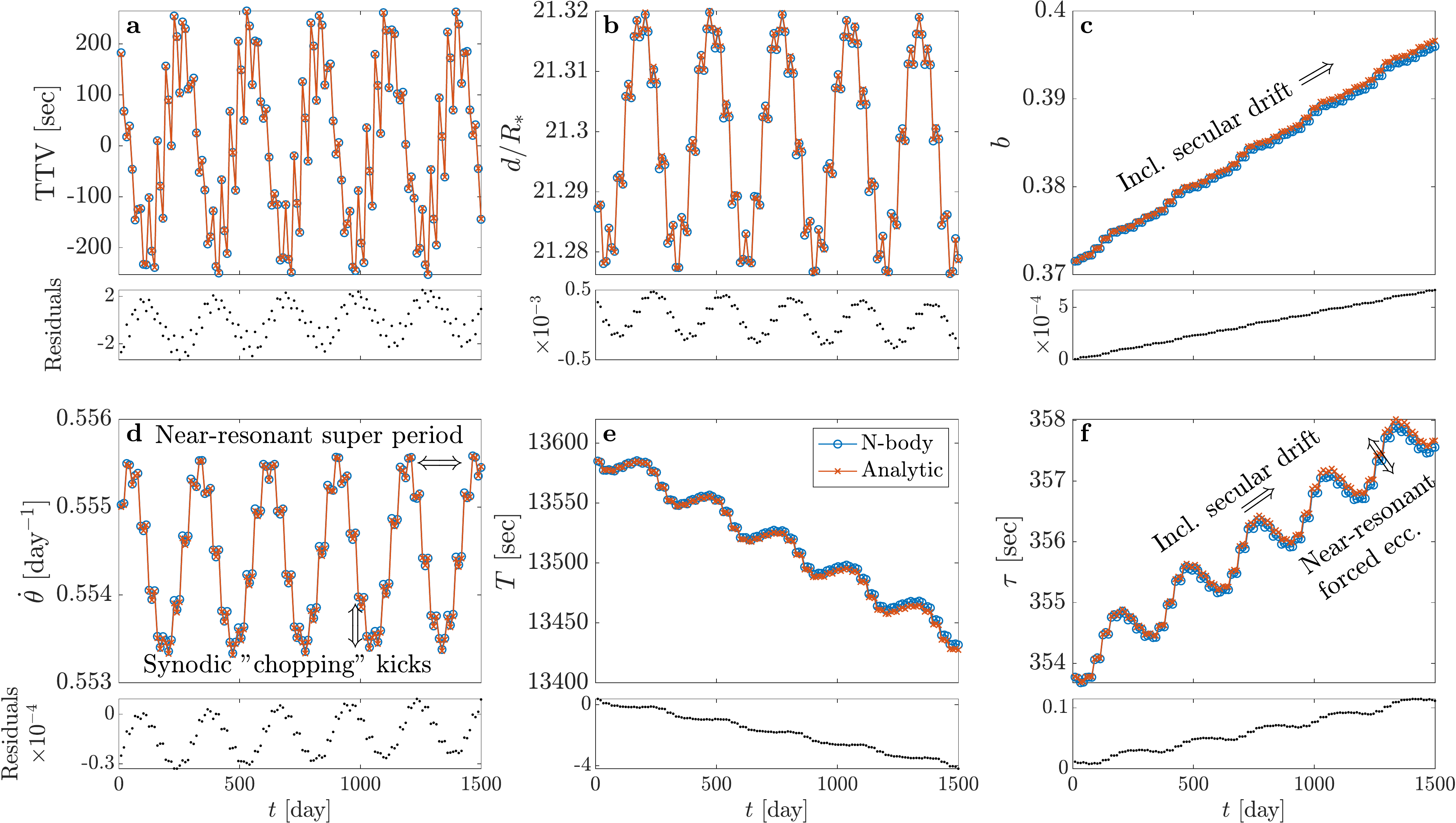}}
    \caption{Individual transit properties of the inner planet of the simulated 2-planets system, showing the ability of \texttt{AnalyticLC} to reproduce a model based on an N-body integration. For each property, the blue circles indicate the results obtained from an N-body integration, while the orange x's are the values obtained from \texttt{AnalyticLC}, with the residuals shown in small black dots. (a) TTV, (b) planet-to-star separation at mid-transit, (c) impact parameter, (d) angular velocity, (e) transit duration, and (f) ingress-egress time.} 
    \label{fig:TransitProps2Planets}
\end{figure}

\begin{figure}
    {
    {\includegraphics[width=1\linewidth]{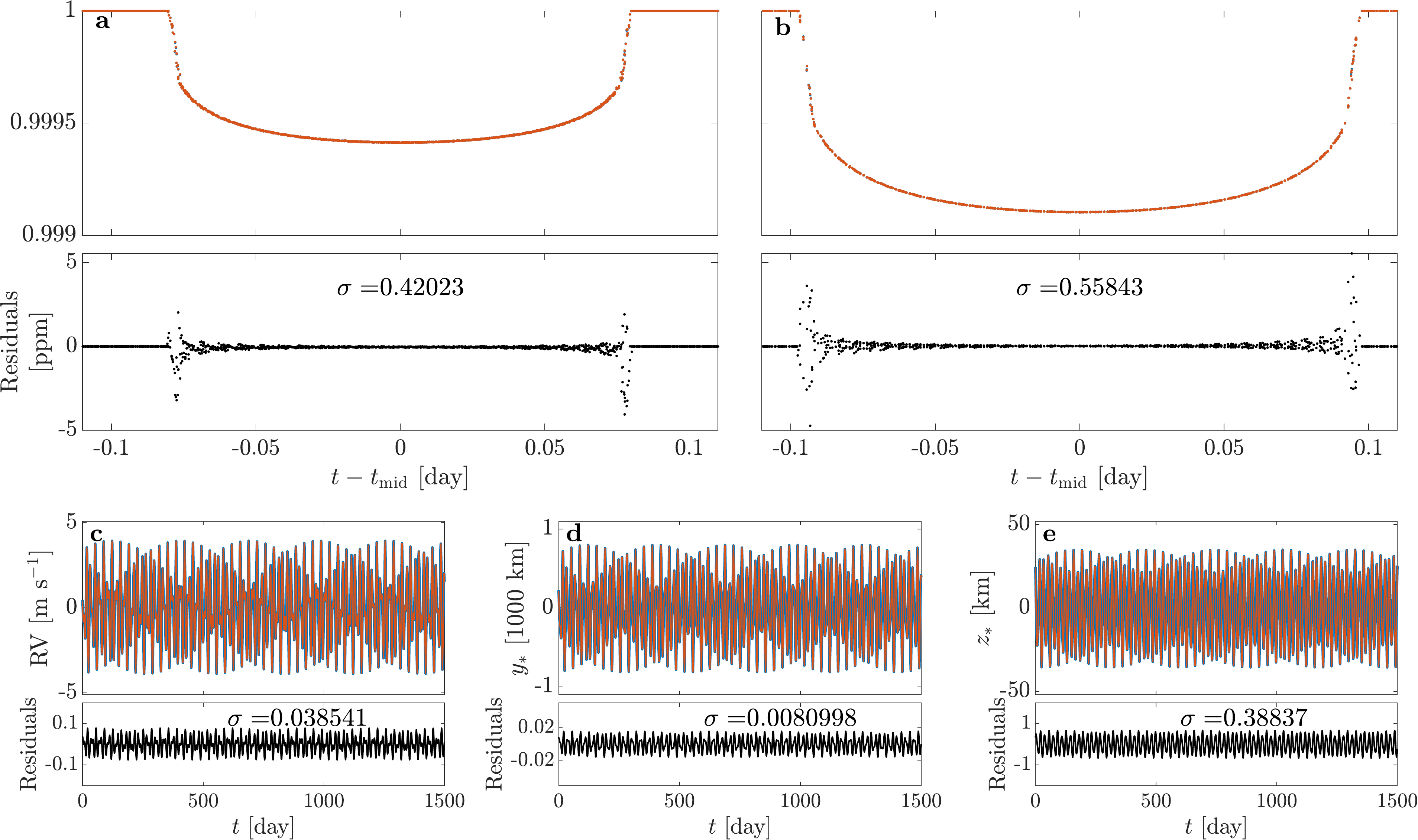}}
    \caption{Model accuracy in terms of light-curve flux, RV and astrometry. As in former figures, results from an N-body integration are shown in blue, while results from \texttt{AnalyticLC} are shown in orange. Residuals are shown in black, with text indicating the standrad deviation $
    \sigma$ of the residuals, which are much smaller than the typical precision of Kepler photometry and of current RV instruments. (a) Folded light curve about times of mid-transit (including TTV) of the inner planet. (b) Folded light curve about times of mid-transit (including TTV) of the outer planet.  (c) RV values. (d,e) Astrometry values in absolute units; for actual data the stellar absolute displacement $y_*,z_*$ would be translated to angular displacement. Note the different scales for $y_*$ and $z_*$; for both cases the residuals are two orders of magnitude smaller than the magnitude of the variations.    \label{fig:FluxTwoPlanets}}
}
\end{figure}

As discussed in \S\ref{sec:GeometryToProperties}, the instantaneous orbital elements values are translated to transit parameters. In Figure~\ref{fig:TransitProps2Planets} we show the accuracy of the transit properties of the inner planet in terms of TTV, planet-star separation, impact parameter, angular velocity, transit duration and ingress-egress time. The TTV shape is composed of a fundamental sine-like oscillation at the super-period frequency, superimposed by a saw-tooth-like "chopping" pattern, due to the gravitational impulses at conjunction \citep{DeckAgol2015}. The standard deviation transit time errors is about 3 seconds, much better than the typical uncertainty on mid-transit times of {Kepler} data, and two orders of magnitude smaller than the TTV semi-amplitude. Together with the other transit properties (planet-to-star separation, impact parameter, angular velocity, transit duration and ingress egress time) \texttt{AnalyticLC}'s accuracy yields a flux error of order 1~ppm (shown in Figure~\ref{fig:FluxTwoPlanets}) - about two orders of magnitude better than the typical {Kepler} precision.
The sinusoidal nature of eccentricity components results from the truncated near-resonant terms of the series expansion. The long-term deviation in the inclination vector motion results from the truncated secular terms of the series expansion.
These parameters are then translatable to direct observables; as shown in the figure 
\texttt{AnalyticLC} provides good accuracy.

\subsection{Comparison with N-body - 3 Planets}

\begin{figure}
    {\includegraphics[width=1\linewidth]{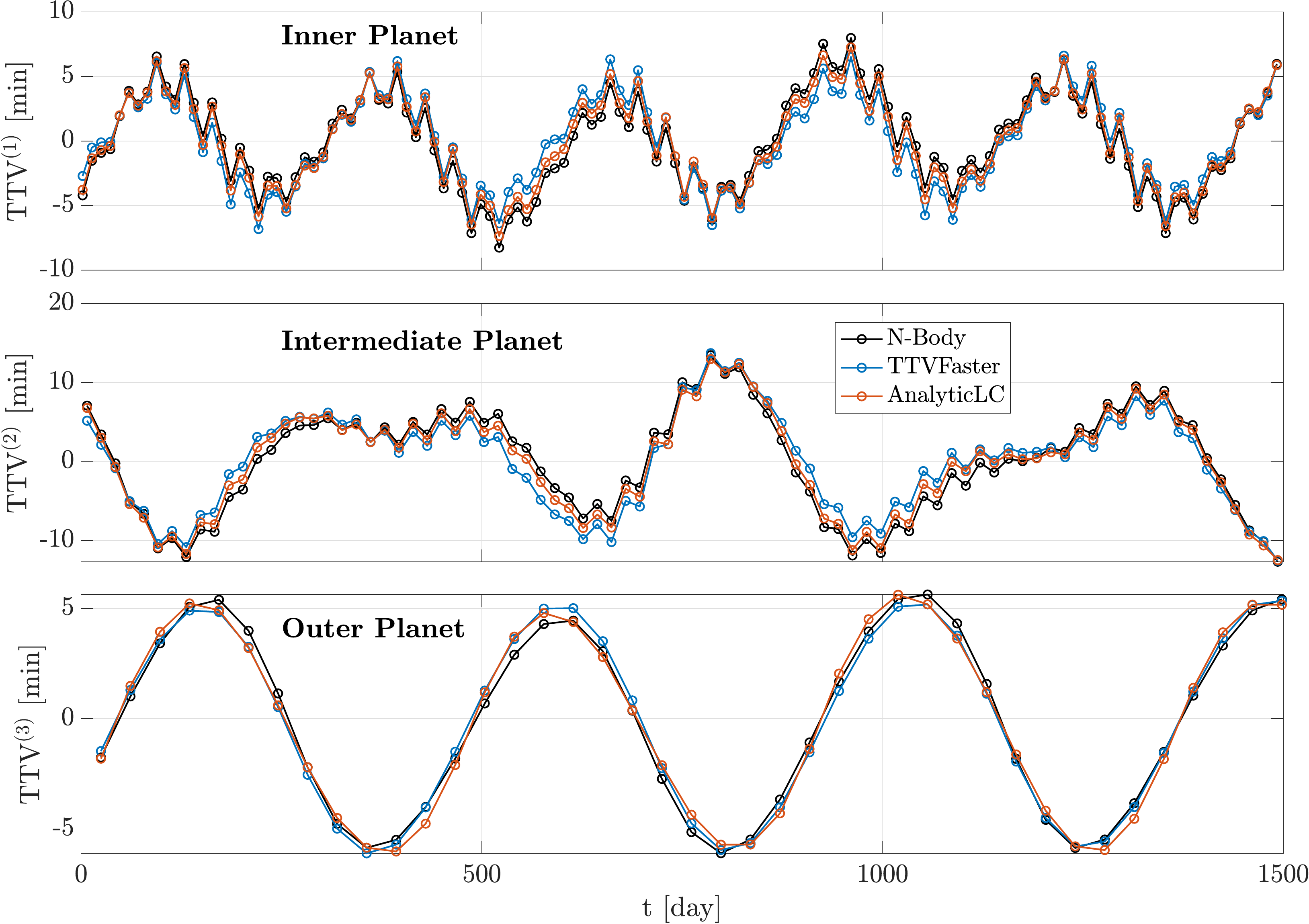}}
    \caption{TTV patterns for a co-planar, small eccentricity, 3-planets system, calculated using different methods: (i) full N-body integration (black) (ii) analytic method of \texttt{TTVFaster} \cite{AgolDeck2016} (blue) (iii) analytic calculation of \texttt{AnalyticLC} (this work), limited to terms of 1st order in eccentricity (red) - matching the \texttt{TTVFaster} analysis. The TTV patterns obtained by both analytic methods qualitatively agree with the full N-body integration; the mismatch between the analytic models and the N-body integration are shown in Figure \ref{fig:ThreePlanetsTmid}.}
    \label{fig:ThreePlanetsTTV}
\end{figure}

\begin{figure}
{\includegraphics[width=.9\linewidth]{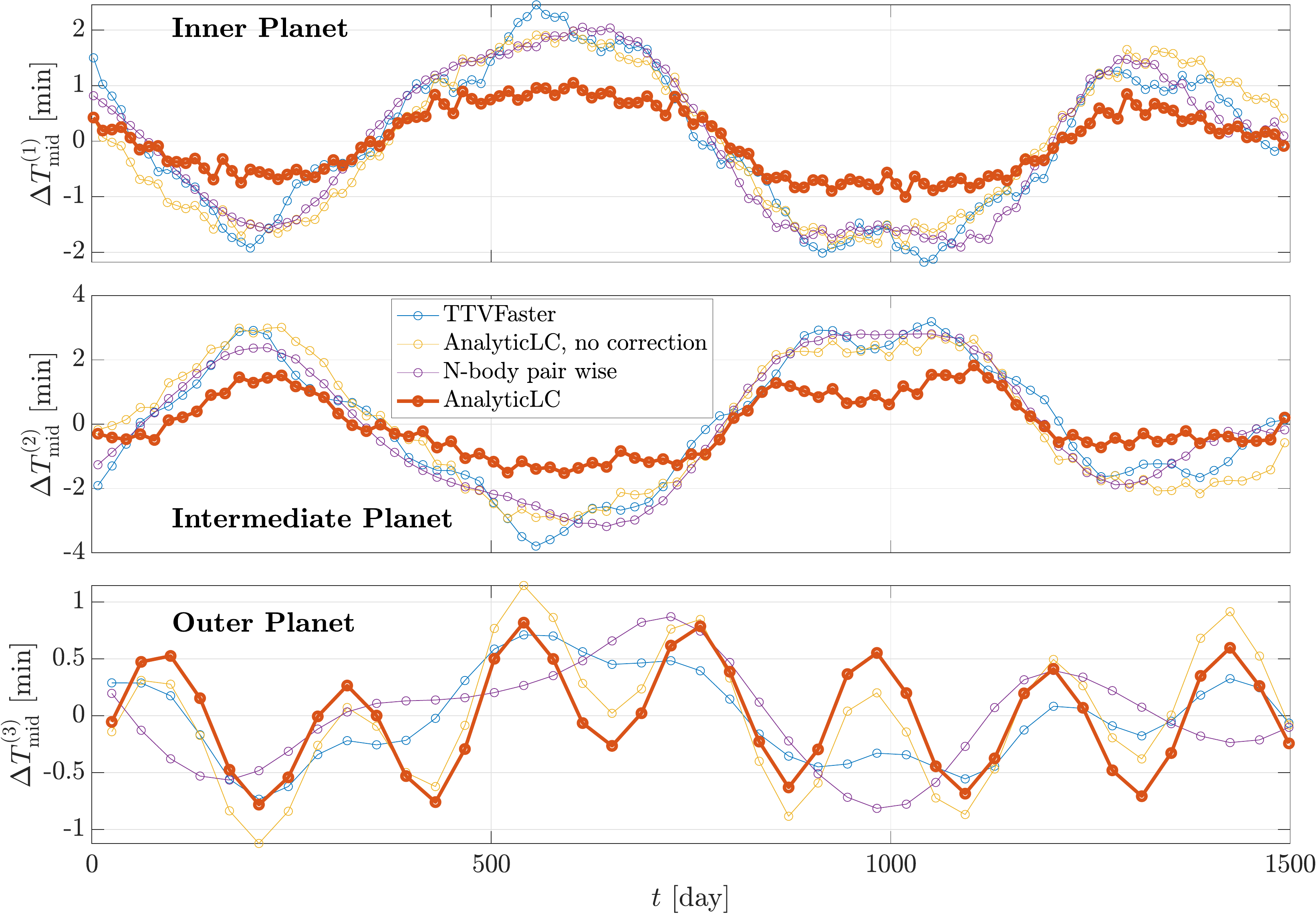}}
    {\caption{Mismatch between the transit times obtained by the analytic methods and the full N-body integration, highlighting the non-additive nature of TTV. The $T_{\rm mid}$ residuals of \texttt{TTVFaster} (blue) and \texttt{AnalyticLC} without the 3-planets correction (yellow) are similar to the residuals obtained by directly adding the TTVs obtained from a pair-wise full N-body integration (purple) - i.e. they are all inaccurate in the same way. This indicates that the residuals are not a result of missing terms in the expansion: the mismatch comes directly from the joint interaction of the 3 planets. The residuals of \texttt{AnalyticLC} for the two innermost planets are reduced by about a half by including the 3-planets interaction correction (orange, highlighted).}
    \label{fig:ThreePlanetsTmid}}
    
\end{figure}

As described above, modeling planet pairs to high accuracy cannot be naively extended to triplets by summing just the TTVs of three pairs. Below we show an example chosen to illustrate this effect.
We model a system of three planets, and calculate the TTVs with numerical (full N-body integration) and analytical methods (\texttt{AnalyticLC} and \texttt{TTVFaster} \citep{AgolDeck2016}) to evaluate this inaccuracy.  Figure~\ref{fig:ThreePlanetsTTV} shows the arising TTV pattern of each planet, calculated by an N-body integration (black), by \texttt{TTVFaster} (blue), and by \texttt{AnalyticLC} (orange). It is visually clear that the analytic methods capture the TTVs of all planets, which is composed of both super-period fundamental TTV \citep{LithwickXieWu2012} and a saw-tooth like "chopping" TTV \citep{DeckAgol2015}.  
We next turn to inspecting the inaccuracy in times-of-mid-transit. Although the large scale TTVs were well-captured by all techniques, upon comparing with the full N-body simulation, structured residuals appear, shown in  Figure~\ref{fig:ThreePlanetsTmid}. The inaccuracy pattern of \texttt{TTVFaster} (blue) is similar to the inaccuracy pattern of \texttt{AnalyticLC} when used without the three-planets term (yellow). In order to prove that these remaining residuals are a direct result of the three planets simultaneous interaction, and not from a  missing term in the pair-wise calculation, we performed a set of pair-wise N-body integrations and summed the TTVs resulting from them (purple); the arising residuals are similar to the analytic methods. Applying \texttt{AnalyticLC} including The 3-planets SMMR terms corrects (some of) this residual TTV pattern (orange, highlighted). We note that  residuals with a similar pattern to the correction itself remain unresolved.

\clearpage

\section{Summary and Future Prospects}\label{sec:Summary}
In this work we presented \texttt{AnalyticLC} - an analytic method and code implementation for light curve/RV/Astrometry modeling. The development of the method was motivated by the benefit of a full light curve modeling over fitting times of mid-transit, which can be used to detect small amplitude TTVs \citep{OfirEtAl2018} and also be used to detect TTVs, which are a key observable for the detection of forces out of the plane. The method assumes that the system is out of resonance, which is the common case in the Kepler population \citep[e.g.][]{FabryckyEtAl2014}. The calculation in resonant terms is up to fourth order in eccentricities and inclinations, sufficiently accurate for systems of up to mild eccentricities, also common in the Kepler population \citep{FabryckyEtAl2014}. For computational efficiency, the calculation can be truncated at lower orders. Beyond the question of calculation speed, such an analytic approach provides insight to the motion of the orbital elements which is difficult to obtain from numerical integration.

We show that 3-planet interactions are important in some cases, in addition to the pairwise interactions in the systems, and make progress towards correctly modeling this effect. We found that these terms correctly predict the morphology of the needed correction but the predicted amplitude of the correction sometimes underestimates the magnitude seen in comparison to full N-body integrations. Because the basic pattern of the residuals is captured by our triple-interaction expression, we are optimistic that including additional terms will further improve the accuracy of the theory\footnote{We offer a reward for the capture of the wanted terms.}.

The computation time of \texttt{AnalyticLC} scales with the number of calculated points - but not with the temporal baseline of the simulation. This advantage becomes increasingly important as data sets begin to span multiple decades (already common in RV, and soon in photometry as well).

In a companion paper, we use this method to fit a model to a subset of the Kepler systems. The method of \texttt{AnalyticLC} can be further used to analyze combined data sets of photometry, RV and astrometry.

Our code is available at \textbf{\url{https://github.com/yair111/AnalyticLC}} along with all necessary files and with a user manual.

\acknowledgments
This study was supported by the Helen Kimmel Center for Planetary Sciences and the Minerva Center for Life Under Extreme Planetary Conditions \#13599 at the Weizmann Institute of Science.

\clearpage

\appendix
\section{orbital elements and transit times derivation}\label{app-A}

\subsection{Equations of Motion}

The mathematical formalism for deriving the orbital elements is based on expanding the disturbing function in powers of $e$ and $I$, and then approximately solving Lagrange's planetary equations. The task of finding the times-of-mid-transit has already been addressed in a similar manner by \cite{HaddenLithwick2016} and \cite{AgolDeck2016}. The derivation of times-of-mid-transit as presented below follows the approach of \cite{HaddenLithwick2016}, adding more terms to the summation and elucidating the role of the different terms in the expansion. Other transit properties are obtained from the instantaneous orbital elements; the expressions are given in $\S$1.
 
The disturbing function is given as a series sum of cosines, with prefactors depending on the ratio $\alpha=a/a'$ multiplied by different powers of the eccentricities and inclinations. Each cosine argument includes a linear combination of the mean longitudes, longitudes of periapse and longitudes of ascending nodes:
\begin{equation}
 \mathcal{R} = Gm'\sum_{j=-\infty}^{j=\infty}S(a,a',e,e',I,I')\cos{\varphi},
\end{equation}
 
where $\varphi$ is a linear combination of the angles $\lambda,\varpi,\Omega,\lambda',\varpi',\Omega'$.
A similar expression gives $\mathcal{R'}$ for the case of an internal perturber. The full derivation of the disturbing function is given in \citet{SSD1999}.

Lagrange's planetary equations are \citep{SSD1999}:
\begin{equation}\label{eq:da_dt}
    \frac{da}{dt} = \frac{2}{na}\frac{\partial{\mathcal{R}}}{\partial\lambda},
\end{equation}
\begin{equation}
    \frac{de}{dt} = -\frac{\sqrt{1-e^2}}{na^2e}(1-\sqrt{1-e^2})\frac{\partial{\mathcal{R}}}{\partial\lambda}-\frac{\sqrt{1-e^2}}{na^2e}\frac{\partial{\mathcal{R}}}{\partial\varpi},
\end{equation}
\begin{equation}
    \frac{d\lambda}{dt} = n\left(1-\frac{3}{2}\frac{\delta a}{a}\right) -\frac{2}{na}\frac{\partial{\mathcal{R}}}{\partial a}+\frac{\sqrt{1-e^2}}{na^2e}(1-\sqrt{1-e^2})\frac{\partial{\mathcal{R}}}{\partial e}+\frac{\tan{(I/2)} }{na^2\sqrt{1-e^2}}\frac{\partial{\mathcal{R}}}{\partial I},
\end{equation}
\begin{equation}
    \frac{d\Omega}{dt} = \frac{1}{na^2\sqrt{1-e^2}\sin{I}}\frac{\partial{\mathcal{R}}}{\partial I},
\end{equation}
\begin{equation}
    \frac{d\varpi}{dt} = \frac{\sqrt{1-e^2}}{na^2e}\frac{\partial{\mathcal{R}}}{\partial e}+\frac{\tan{(I/2)}}{na^2e\sqrt{1-e^2}}\frac{\partial{\mathcal{R}}}{\partial I},
\end{equation}
\begin{equation}
    \frac{d I}{dt} = -\frac{\tan{(I/2)}}{na^2\sqrt{1-e^2}}\left(\frac{\partial{\mathcal{R}}}{\partial \lambda}+\frac{\partial{\mathcal{R}}}{\partial \varpi}\right)-\frac{1}{na^2\sqrt{1-e^2}\sin{I}}\frac{\partial{\mathcal{R}}}{\partial\Omega}.
\end{equation}

We solve them in two steps: 

(i) Solving the secular motion of the eccentricity and inclination vectors \citep[chapter 7]{SSD1999} (using a second-order expansion of the disturbing function; the next order correction is 4th power). 

(ii) Solving the near-resonant interactions by approximating that the right-hand-side is time dependent only through the mean longitudes. 

A subtlety is noted here regarding the calculation of the derivatives with respect to the semi-major axes. For these derivatives, we assume that the perturber's semi-major axis is held constant, and then represent the derivative with respect to the planetary semi-major axis by the derivative with respect to $\alpha=a/a'$, using the chain rule: $\frac{\partial}{\partial a}=\frac{1}{a'}\frac{\partial}{\partial\alpha}$ and $\frac{\partial}{\partial a'}=-\frac{a}{a'^2}\frac{\partial}{\partial\alpha}$. This enables the calculation of the derivatives of the Laplace coefficients, which are functions of $\alpha$.

\subsection{General Approximate Solution}
In this section we describe the second step of the method: solution of Lagrange's planetary equations (the first step, calculating the secular motion, is  described in \citet[chapter 7]{SSD1999}). We make use of the fact that the equations are linear in $\mathcal{R}$, and hence sum the effects of individual terms in the general form.
Given the orbital elements of the inner companion $a,e,\varpi,I,\Omega,\lambda$ and their counterparts denoted with a prime corresponding to the outer companion, we define the following quantities:
\begin{equation}
    C_{jk} = e^Ae'^{A'}s^Bs'^{B'}\cos{(j\lambda'+(k-j)\lambda-C\varpi-C'\varpi'-D\Omega-D'\Omega')}
\end{equation}
and
\begin{equation}
    S_{jk} = e^Ae'^{A'}s^Bs'^{B'}\sin{(j\lambda'+(k-j)\lambda-C\varpi-C'\varpi'-D\Omega-D'\Omega')},
\end{equation}
with $s=\sin(I/2)$ and $s'=\sin(I'/2)$ \citep{SSD1999}.
The disturbing function of the inner planet, then, can be given by
\begin{equation}
    \mathcal{R}_{jk} = \frac{Gm'}{a'}(f+f_E)C_{jk},
\end{equation}
where $f$ and $f_E$ are functions that depend on $\alpha=a/a'$ and on the specific values of $j$ and $k$, but not on the orbital elements. The separation to two parts comes from the distinction between the direct part and the indirect part of the disturbing function for an external perturber that exist for specific cosine arguments.

The disturbing function for the outer planet is given in a similar form:
\begin{equation}
    \mathcal{R'}_{jk} = \frac{Gm}{a'}(f+f_I)C_{jk} = \frac{Gm}{a}(\alpha f+\alpha f_I)C_{jk}.
\end{equation}
The factor $f_I$ contains the additional terms that sometimes exist, that arise from the indirect part for an internal perturber.

If we assume that $C_{jk}$ and $S_{jk}$ depend on time only through the mean longitudes, we can establish the following relations:
\begin{equation} \label{eq:int_Cjk}
    \int C_{jk}dt = \frac{1}{n_{jk}}S_{jk}
\end{equation}
and
\begin{equation} \label{eq:int_Sjk}
    \int S_{jk}dt = -\frac{1}{n_{jk}}C_{jk},
\end{equation}
where 
\begin{equation}
    n_{jk} = jn'+(k-j)n
\end{equation}
is the mean motion of the longitude of conjunction.

We note here again, as mentioned previously, that one must assume that the longitudes of conjunction circulate rather than librate for this calculation to hold.

Next, we express the derivatives of the single disturbing function term using $C_{jk}, S_{jk}$:
\begin{equation}
    \frac{\partial\mathcal{R}_{jk}}{\partial\lambda} = (j-k)S_{jk},
\end{equation}
\begin{equation}
    \frac{\partial\mathcal{R}_{jk}}{\partial e} = \frac{A}{e}C_{jk},
\end{equation}
\begin{equation}
    \frac{\partial\mathcal{R}_{jk}}{\partial\varpi} = CS_{jk},
\end{equation}
\begin{equation}
    \frac{\partial\mathcal{R}_{jk}}{\partial I} = \frac{B\cot{(I/2)}}{2}C_{jk},
\end{equation}
\begin{equation}
    \frac{\partial\mathcal{R}_{jk}}{\partial\Omega} = DS_{jk},
\end{equation}
and
\begin{equation}
    \frac{\partial\mathcal{R}_{jk}}{\partial a} = \frac{1}{a'}\frac{\frac{\partial f}{\partial\alpha}+\frac{\partial f_E}{\partial\alpha}}{f+f_E}C_{jk},
\end{equation}
where for the equations for $\frac{\partial\mathcal{\langle R \rangle}}{\partial I}$ and $\frac{\partial\mathcal{\langle R \rangle}}{\partial a}$ we used the chain rule, as following:
\begin{equation}
    \frac{\partial\mathcal{R}_{jk}}{\partial I}=\frac{\partial\mathcal{R}_{jk}}{\partial s}\frac{\partial s}{\partial I}
\end{equation}
and
\begin{equation}
    \frac{\partial\mathcal{R}_{jk}}{\partial a}=\frac{\partial\mathcal{R}_{jk}}{\partial \alpha}\frac{\partial \alpha}{\partial a}.
\end{equation}

Now we substitute these derivatives in Lagrange's planetary equations, and integrate them in time using equations \ref{eq:int_Cjk} and \ref{eq:int_Sjk}. This yields the {\it variations} in the orbital elements. In order to put the equations in dimensionless form, we make use of Kepler's law, and define the relative masses
\begin{equation}
    \mu = \frac{m}{m_*}
\end{equation}
and
\begin{equation}
    \mu' = \frac{m'}{m_*},
\end{equation}
where $m_*$ is the stellar mass.





The obtained expressions are:
\begin{equation}\label{eq:da_oa}
    \frac{\delta a}{a} = 2\frac{\mu'}{1+\mu}\alpha(f+f_E)(k-j)\frac{n}{n_{jk}}C_{jk},
\end{equation}
\begin{eqnarray}
    \delta\lambda &= & 3\frac{\mu'}{1+\mu}\alpha(f+f_E)(j-k)\frac{n^2}{n_{jk}^2}S_{jk} \nonumber \\
    &- & 2\frac{\mu'}{1+\mu}\alpha^2(\frac{\partial f}{\partial \alpha}+\frac{\partial f_E}{\partial \alpha})\frac{n}{n_{jk}}S_{jk} \nonumber \\
    &+ &\frac{\mu'}{1+\mu}\alpha(f+f_E)A\frac{\sqrt{1-e^2}(1-\sqrt{1-e^2})}{e^2}\frac{n}{n_{jk}}S_{jk} \nonumber \\
    &+ &\frac{\mu'}{1+\mu}\alpha(f+f_E)\frac{B}{2}\frac{1}{\sqrt{1-e^2}}\frac{n}{n_{jk}}S_{jk},
\end{eqnarray}
\begin{eqnarray}
        \delta e &= & \frac{\mu'}{1+\mu}\alpha(f+f_E)(j-k)\frac{\sqrt{1-e^2}(1-\sqrt{1-e^2})}{e}\frac{n}{n_{jk}}C_{jk} \nonumber \\
        &+ & \frac{\mu'}{1+\mu}\alpha(f+f_E)C\frac{\sqrt{1-e^2}}{e}\frac{n}{n_{jk}}C_{jk}
\end{eqnarray}
\begin{eqnarray}
        \delta \varpi &= & \frac{\mu'}{1+\mu}\alpha(f+f_E)A\frac{\sqrt{1-e^2}}{e^2}\frac{n}{n_{jk}}S_{jk} \nonumber \\
        &+ & \frac{\mu'}{1+\mu}\alpha(f+f_E)\frac{B}{2}\frac{1}{\sqrt{1-e^2}}\frac{n}{n_{jk}}S_{jk}
\end{eqnarray}
\begin{eqnarray}
        \delta I &= & \frac{\mu'}{1+\mu}\alpha(f+f_E)(j-k+C)\frac{\tan{(I/2)}}{\sqrt{1-e^2}}\frac{n}{n_{jk}}C_{jk} \nonumber \\
        &+ & \frac{\mu'}{1+\mu}\alpha(f+f_E)D\frac{1}{\sqrt{1-e^2}\sin{I}}\frac{n}{n_{jk}}C_{jk}
\end{eqnarray}
\begin{eqnarray}
        \delta\Omega &=& \frac{\mu'}{1+\mu}\alpha(f+f_E)\frac{B}{2}\frac{\cot{(I/2)}}{\sqrt{1-e^2}\sin{I}}\frac{n}{n_{jk}}S_{jk}.
\end{eqnarray}

A similar derivation for the external companion yields:

\begin{equation}
        \frac{\delta a'}{a'} = 2\frac{\mu}{1+\mu'}(f+f_I)j\frac{n'}{n_{jk}}C_{jk}
\end{equation}
\begin{eqnarray}
        \delta \lambda' &=& -3\frac{\mu}{1+\mu'}(f+f_I)j\frac{n'^2}{n_{jk}^2}S_{jk} \nonumber \\
        &+ &2\frac{\mu}{1+\mu'}\left(f+f_I+\alpha \frac{\partial f}{\partial\alpha}+\alpha\frac{\partial f_I}{\partial\alpha} \right)\frac{n'}{n_{jk}}S_{jk} \nonumber \\
        &+ & \frac{\mu}{1+\mu'}(f+f_I)A'\frac{\sqrt{1-e'^2}(1-\sqrt{1-e'^2})}{e'^2}\frac{n'}{n_{jk}}S_{jk} \nonumber \\
        &+ & \frac{\mu}{1+\mu'}(f+f_I)\frac{B'}{2}\frac{1}{\sqrt{1-e'^2}}\frac{n'}{n_{jk}}S_{jk}
\end{eqnarray}
\begin{eqnarray}
        \delta e' &=& -\frac{\mu}{1+\mu'}(f+f_I)j\frac{\sqrt{1-e'^2}(1-\sqrt{1-e'^2})}{e'}\frac{n'}{n_{jk}}C_{jk} \nonumber \\
        &+& \frac{\mu}{1+\mu'}(f+f_I)C'\frac{\sqrt{1-e'^2}}{e'}\frac{n'}{n_{jk}}C_{jk}
\end{eqnarray}
\begin{eqnarray}
        \delta \varpi' &=& \frac{\mu}{1+\mu'}(f+f_I)A'\frac{\sqrt{1-e'^2}}{e'^2}\frac{n'}{n_{jk}}S_{jk} \nonumber \\
        &+& \frac{\mu}{1+\mu'}(f+f_I)\frac{B'}{2}\frac{1}{\sqrt{1-e'^2}}\frac{n'}{n_{jk}}S_{jk}
\end{eqnarray}
\begin{eqnarray}
        \delta I' &=& \frac{\mu}{1+\mu'}(f+f_I)(C'-j)\frac{\tan{(I'/2)}}{\sqrt{1-e'^2}}\frac{n'}{n_{jk}}C_{jk} \nonumber \\
        &+&\frac{\mu}{1+\mu'}(f+f_I)D'\frac{1}{\sqrt{1-e'^2}\sin{I'}}\frac{n'}{n_{jk}}C_{jk}
\end{eqnarray}
\begin{eqnarray}
        \delta \Omega' = \frac{\mu}{1+\mu'}(f+f_I)\frac{B'}{2}\frac{\cot{(I'/2)}}{\sqrt{1-e'^2}\sin{I'}}\frac{n'}{n_{jk}}S_{jk}.
\end{eqnarray}

The variations in the complex eccentricity vector $z=ee^{i\varpi}$ and the complex inclination vector $u=Ie^{i\Omega}$ are given by
\begin{equation}
    \delta z = z \left(\frac{\delta e}{e}+i\delta\varpi\right)
\end{equation}
and
\begin{equation}
    \delta u = u\left(\frac{\delta I}{I}+i\delta\Omega\right),
\end{equation}
with similar expressions for the variations of the external planet complex vectors $z', u'$.

\subsection{Mid-transit times}
The translation of orbital elements variations to transit times has been done in former works for the coplanar case to first order in eccentricity \citep{LithwickXieWu2012,DeckAgol2016} and second order in eccentricity \citep{HaddenLithwick2016}. The calculation here follows the same principle, though we expand the true longitude to the fourth power in eccentricity. For our convention of the $x$ axis pointing from the stellar center to the observer, transits occurs at true longitude $\theta=0$. The transit timing variations are given by transforming the angular variations to timing variations.

To fourth order in eccentricity, the true anomaly is given by \citep[eq. 2.88]{SSD1999},
\begin{equation}
    f = M+2e\sin{M}+\frac{5}{4}e^2\sin{2M}+e^3\left(\frac{13}{12}\sin{3M}-\frac{1}{4}\sin{M}\right)+e^4\left(\frac{103}{96}\sin{4M}-\frac{11}{24}\sin{2M}\right).
\end{equation}

Using the relations $f+\varpi=\theta$ and $\lambda=\varpi+M$ we obtain
\begin{eqnarray}
    \theta &= & \lambda + 2e\sin{(\lambda-\varpi)}+\frac{5}{4}e^2\sin{(2(\lambda-\varpi))} \nonumber \\ &+ & e^3\left(\frac{13}{12}\sin{(3(\lambda-\varpi))}-\frac{1}{4}\sin{(\lambda-\varpi)}\right) \nonumber \\
    &+ & e^4\left(\frac{103}{96}\sin{(4(\lambda-\varpi))}-\frac{11}{24}\sin{(2(\lambda-\varpi))}\right).
\end{eqnarray}

Most of the terms in the expression for $\theta$ have the same power of $e$ as the pre-factor of $M$. Therefore, we express the general form of these terms using the complex eccentricity $z=e\,e^{i\varpi}$ and the relation $M=\lambda-\varpi$ as follows:
\begin{equation}
    e^k\sin{(kM)} = \Re\left(\frac{(e^{i\lambda}z^*)^k}{i}\right)
\end{equation}
for any $k$.

Taking the differential with respect to the variables $\lambda, z$ yields

\begin{equation}
    \delta(e^k\sin{(kM)}) = k\Re\left(e^{ik\lambda}\left(z^{*k}\delta\lambda+\frac{z^{*k-1}\delta z^*}{i}\right)\right).
\end{equation}

There are two terms in the expression for $\theta$ for which the power of $e$ does not match the pre-factor of $M$; for expressing their differential we make use of the identity $e^2=zz^*$, and get:
\begin{equation}
    \delta(e^3\sin{M})=(z^*\delta z+z\delta z^*)\Re\left(\frac{e^{i\lambda}z^*}{i}\right)+zz^*\Re\left(e^{i\lambda}\left(z^*\delta\lambda+\frac{\delta z^*}{i}\right)\right)
\end{equation}
and
\begin{equation}
    \delta(e^4\sin{2M})=(z^*\delta z+z\delta z^*)\Re\left(\frac{(e^{i\lambda}z^*)^2}{i}\right)+2zz^*\Re\left(e^{2i\lambda}\left(z^{*2}\delta\lambda+\frac{z^*\delta z^*}{i}\right)\right).
\end{equation}

Summing all these differentials yields an expression for $\delta\theta$ as a function of the complex eccentricity and the mean longitudes:
\begin{eqnarray}
    \delta\theta &= & \delta\lambda +2\Re\left(e^{i\lambda}\left(z^*\delta\lambda+\frac{\delta z^*}{i}\right)\right) \nonumber \\ 
    &+ &\frac{5}{4}2\Re\left(e^{2i\lambda}\left(z^{*2}\delta\lambda+\frac{z^{*}\delta z^*}{i}\right)\right) \nonumber \\
    &+ &\frac{13}{12}3\Re\left(e^{3i\lambda}\left(z^{*3}\delta\lambda+\frac{z^{*2}\delta z^*}{i}\right)\right) \nonumber \\
    &- &\frac{1}{4}\left((z^*\delta z+z\delta z^*)\Re\left(\frac{e^{i\lambda}z^*}{i}\right)+zz^*\Re\left(e^{i\lambda}\left(z^*\delta\lambda+\frac{\delta z^*}{i}\right)\right)\right) \nonumber \\
    &+ & \frac{103}{96}4\Re\left(e^{4i\lambda}\left(z^{*4}\delta\lambda+\frac{z^{*3}\delta z^*}{i}\right)\right) \nonumber \\
    &- & \frac{11}{24}\left((z^*\delta z+z\delta z^*)\Re\left(\frac{(e^{i\lambda}z^*)^2}{i}\right)+2zz^*\Re\left(e^{2i\lambda}\left(z^{*2}\delta\lambda+\frac{z^*\delta z^*}{i}\right)\right)\right).
\end{eqnarray}

In fact, for the the sake of calculating the TTV, we are interested only in this expression for $\theta=0$ (and $\lambda \approx 0)$.

Relating $\delta\theta$ to time is done from the expression for the angular velocity:
\begin{equation}
    \dot{\theta} = n\frac{(1+e\cos{f})^2}{(1-e^2)^{3/2}}.
\end{equation}

If we are interested only in the values for $\theta=0$ ({\it i.e.} when transit occurs) we get $f=-\varpi$ and hence
\begin{equation}
    \dot{\theta}(\theta=0) = n\frac{(1+e\cos{\varpi})^2}{(1-e^2)^{3/2}}.
\end{equation}

The TTV, denoted by $\delta t$, is the translation of the angular shift to temporal shift. Larger longitude implies an earlier transit and negative $\delta t$. The explicit expression for the TTV is hence
\begin{equation}
    \delta t = -\delta\theta \frac{P}{2\pi}\frac{(1-e^2)^{3/2}}{(1+\Re(z))^2},
\end{equation}
with $\delta\theta$ defined above.

\subsection{TTVs of 3 planets in a near-resonant chain}
In this section we derive TTV patterns which are second order in mass that arise from the cross-interaction between two pairs. These TTV patterns make the total TTV deviate from the sum of the pairwise interactions.
We analyze a system of three co-planar planets, assuming that the innermost planet and the intermediate planet are near the $j:j-1$ MMR, and that the intermediate planet and the outermost planet are near the $k:k-1$ MMR. We neglect the interaction between the innermost and outermost planet to first order in eccentricity, assuming that they are not near any first order MMR - this is justified if the system is not too packed.
We use the same variables as described above, but here we also add the double prime to account for the variables relating to the outermost planet.
We refer here only to the terms in the disturbing function related to the $j:j-1$ and $k:k-1$ resonance (as in \citet{LithwickXieWu2012}).
The averaged disturbing function in this case is:
\begin{equation}
\mathcal{\langle R \rangle} = \frac{Gm'}{a'}( f_{27}(\alpha_{12})e\cos{(j\lambda'+(1-j)\lambda-\varpi)}+(f_{31}(\alpha_{12})-2\alpha_{12}\delta_{j,2})e'\cos{(j\lambda'+(1-j)\lambda-\varpi')}),
\end{equation}
\begin{eqnarray}
\mathcal{\langle R' \rangle} &= & \frac{Gm}{a'}\left( f_{27}(\alpha_{12})e\cos{(j\lambda'+(1-j)\lambda-\varpi)}+\left(f_{31}(\alpha_{12})-\frac{\delta_{j,2}}{2\alpha_{12}^2}\right)e'\cos{(j\lambda'+(1-j)\lambda-\varpi')}\right) \nonumber \\
&+ &\frac{Gm''}{a''}(f_{27}(\alpha_{23})e'\cos{(k\lambda''+(1-k)\lambda'-\varpi')}+(f_{31}(\alpha_{23})-2\alpha_{23}\delta_{k,2})e''\cos{(k\lambda''+(1-k)\lambda'-\varpi'')}), \nonumber \\
\end{eqnarray}
and
\begin{equation}
\mathcal{\langle R'' \rangle} = \frac{Gm'}{a''}
\left(f_{27}(\alpha_{23})e'\cos{(k\lambda''+(1-k)\lambda'-\varpi')}+\left(f_{31}(\alpha_{23})-\frac{\delta_{k,2}}{2\alpha_{23}^2}\right)e''\cos{(k\lambda''+(1-k)\lambda'-\varpi'')}\right).
\end{equation}

Solving these equations for the motion of the eccentricity vector gives the sum of the motions for each pair interaction, as described by \citet{LithwickXieWu2012}:
\begin{equation}
    z = z^{(\rm free)}+\frac{Gm'}{na^2a''n^j}f_{27}(\alpha_{12})e^{i\lambda^j},
\end{equation}
\begin{equation}
    z' = z'^{(\rm free)} +\frac{Gm}{n'a'^3n^j}\left(f_{31}(\alpha_{12})-\frac{\delta_{j,2}}{2\alpha_{12}^2}\right)e^{i\lambda^j}+\frac{Gm''}{n'a'^2a''n^k}f_{27}(\alpha_{23})e^{i\lambda^k},
\end{equation}
and
\begin{equation}
    z'' = z''^{(\rm free)} +\frac{Gm'}{n''a''^3n^k}\left(f_{31}(\alpha_{23})-\frac{\delta_{k,2}}{2\alpha_{23}^2}\right)e^{i\lambda^k}.
\end{equation}

Heuristically, each first order near-MMR interactions creates a forced eccentricity that draws a circle about the free eccentricity in the eccentricity plane, as previously described by \citet{LithwickXieWu2012}. We also derive the TTV in a similar manner: translating the variations in the semi-major axis to variations in the mean motion.

The variations in the semi-major axis are given by Eq.~\ref{eq:da_dt},
with analogous equations for the evolution of $a',a''$.

Deriving the disturbing functions with respect to the mean longitudes yields
\begin{equation}
    \frac{\partial\mathcal{R}}{\partial\lambda} = 
    \frac{Gm'}{a'}(j-1)(f_{27}(\alpha_{12})e\sin{(\lambda^j-\varpi)}+(f_{31}(\alpha_{12})-2\delta_{j,2}\alpha_{12})e'\sin{(\lambda^j-\varpi')}),
\end{equation}
\begin{eqnarray}
    \frac{\partial\mathcal{R'}}{\partial\lambda'} =  &- &\frac{Gm}{a'}j\left(f_{27}(\alpha_{12})e\sin{(\lambda^j-\varpi)}+\left(f_{31}(\alpha_{12})-\frac{\delta_{j,2}}{2\alpha_{12}^2}\right)e'\sin{(\lambda^j-\varpi')}\right) \nonumber \\
    &+ &\frac{Gm''}{a''}(k-1)(f_{27}(\alpha_{23})e'\sin{(\lambda^k-\varpi')}+(f_{31}(\alpha_{23})-2\delta_{k,2}\alpha_{23})e''\sin{(\lambda^k-\varpi'')}),
\end{eqnarray}
and
\begin{equation}
    \frac{\partial\mathcal{R''}}{\partial\lambda''} = -\frac{Gm'}{a''}k\left(f_{27}(\alpha_{23})e'\sin{(\lambda^k-\varpi')}+\left(f_{31}(\alpha_{23})-\frac{\delta_{k,2}}{2\alpha_{23}^2}\right)e''\sin{(\lambda^k-\varpi'')}\right),
\end{equation}
where $\lambda^j=\lambda'+(1-j)\lambda$ and $\lambda^k=\lambda''+(1-k)\lambda'$, following the notations of \citet{LithwickXieWu2012}.

The variations in the mean longitudes are given approximately by
\begin{equation}
    \frac{d\lambda}{dt}\approx n\left(1-\frac{3}{2}\frac{\delta a}{a}\right),
\end{equation}
where $\delta a$ is the variations in $a$ about its mean value; other terms also appear, but they would have a larger denominator as will be evident soon and hence we neglect them here. Hence, to obtain the variations in $\lambda$ which cause the second-order-in-mass TTVs, we  integrate to get
\begin{equation}
    \delta\lambda \approx -\frac{3}{2}\frac{n}{a}\int\delta a\, dt,
\end{equation}
where $\delta a$ is given by eq-\ref{eq:da_oa}.

This expression is written as an indefinite integral without the two constants of integration, because they are absorbed into the definition of the linear ephemeris parameters (orbital period and reference time).

The derivation for $\delta\lambda', \delta\lambda''$ is analogous.

Let us now inspect the expression for $\partial\mathcal{R}/\partial\lambda$. If we treat $e,e'$ as constant in time, we simply obtain the sum of two near-MMR interactions of \citet{LithwickXieWu2012}. However, we are interested in the cross interaction between the two pairs: plugging in the solution of $z'$ using $e'=z'e^{-i\varpi'}$, we get
\begin{eqnarray}
    \frac{\partial\mathcal{R}}{\partial\lambda} &= &
    \frac{Gm'}{a'}(j-1)(f_{27}(\alpha_{12})e\sin{(\lambda^j-\varpi)}+(f_{31}(\alpha_{12})-2\delta_{j,2}\alpha_{12}) \nonumber \\ 
    & &\left(z'^{(\rm free)}e^{-i\varpi'}+\frac{Gm}{n'a'^3n^j}\left(f_{31}(\alpha_{12})-\frac{\delta_{j,2}}{2\alpha_{12}^2}\right)e^{i(\lambda^j-\varpi')}+\frac{Gm''}{n'a'^2a''n^k}f_{27}(\alpha_{23})e^{i(\lambda^k-\varpi')}\right)\sin{(\lambda^j-\varpi')}). \nonumber \\
\end{eqnarray}

We take the real part of both sides and use the angle-sum identity to integrate this expression in time, yielding a solution for $\delta\lambda$. The terms discussed here are second order in mass, and hence will be significant only if they attain a small denominator due to the integration in time. Therefore, we treat only the second order denominator arising from the double integration and neglect other terms in the equation for $d\lambda/dt$.

Finally, the integration yields the cross term, for which we use the superscript (2-3) to emphasize that this is the cross term arising from the $m'm''$ term in mass (the interaction between planets 2 and 3):
\begin{eqnarray}
    \delta\lambda^{(2-3)}&= & -\frac{3}{2}\frac{Gm'Gm''}{n'a^2a'^3a''}\frac{j-1}{n^k}f_{27}(\alpha_{23})(f_{31}(\alpha_{12})-2\delta_{j,2}\alpha_{12}) \nonumber \\
  & & \left(\frac{\sin{(\lambda^k+\lambda^j-2\varpi')}}{(n^k+n^j)^2}-\frac{\sin{(\lambda^k-\lambda^j)}}{(n^k-n^j)^2}\right)
\end{eqnarray}.
using $\delta t\approx -P\, \delta\lambda/2\pi$, and using Kepler's law $n^2a^3=G(m_*+m), n'^2a'^3=G(m_*+m')$ we get
\begin{eqnarray}
    \delta t^{(2-3)} &= & \frac{P}{2\pi}\frac{3}{2}\frac{m'm''}{(m_*+m)(m_*+m')}n'n^2\alpha_{12}\alpha_{23}\frac{j-1}{n^k}f_{27}(\alpha_{23})(f_{31}(\alpha_{12})-2\delta_{j,2}\alpha_{12})  \nonumber \\
  & & \left(\frac{\sin{(\lambda^k+\lambda^j-2\varpi')}}{(n^k+n^j)^2}-\frac{\sin{(\lambda^k-\lambda^j)}}{(n^k-n^j)^2}\right).
\end{eqnarray}

In a similar manner, we obtain
\begin{eqnarray}
    \delta t'^{(1-3)} &= & \frac{P'}{2\pi}\frac{3}{2}\frac{mm''}{(m_*+m')^2}n'^3\alpha_{23}f_{27}(\alpha_{23})\left(f_{31}(\alpha_{12})-\frac{\delta_{j,2}}{2\alpha_{12}^2}\right) \nonumber \\
  & &  \left(\left(\frac{j}{n^k}+\frac{1-k}{n^j}\right)\frac{\sin{(\lambda^j+\lambda^k-2\varpi')}}{(n^j+n^k)^2}+\left(\frac{j}{n^k}-\frac{1-k}{n^j}\right)\frac{\sin{(\lambda^j-\lambda^k)}}{(n^j-n^k)^2}\right),
\end{eqnarray}
and
\begin{eqnarray}
    \delta t''^{(1-2)} &=&  \frac{P''}{2\pi}\frac{3}{2}\frac{mm'}{(m_*+m)(m_*+m'')}n'n''^2\frac{k}{n^j}f_{27}(\alpha_{23})\left(f_{31}(\alpha_{12})-\frac{\delta_{j,2}}{2\alpha_{12}^2}\right) \nonumber \\ 
    & & \left(\frac{\sin{(\lambda^j+\lambda^k-2\varpi')}}{(n^j+n^k)^2}-\frac{\sin{(\lambda^j-\lambda^k)}}{(n^j-n^k)^2}\right).
\end{eqnarray}

\section{Implementation Considerations}\label{app-B}

The code was implemented in MATLAB, with modular structure and embedded documentation. It takes advantage of vectorized operations where practical.

For computational efficiency, two supporting mechanisms were implemented. The first, is a precalculated table of the Laplace coefficients and their derivatives, which avoids the need for a direct calculation of the integral at each function evaluation. The table is constructed once and the values used many times during the calculation. 

The second mechanism is a look-up table for the Mandel-Agol function values \citep{MandelAgol2002}. The Mandel-Agol model requires four inputs: planet-to-star radii ratio, sky-projected distance in stellar radii and two limb-darkening coefficients. Since in practice, the limb-darkening coefficients are given, then we construct a look-up table for a grid of planet radius and separation values, and perform linear interpolation within the grid. For a grid of 100 planetary radius values and 500 planet-to-star distance values, The typical accuracy cost in the relative flux is at the order of $10^{-7}$, about three orders of magnitude smaller than {\it $Kepler$}'s typical long-cadence error.

We performed timing tests of the code by running a large number of simulations. The calculation time depends on various parameters, such as the number of transits, the number of planets, the order of the expansion etc. To get an idea of the code's performance, we compared the running time of \texttt{AnalyticLC} to the running time of \texttt{TTVFast} for a 2-planets system with orbital periods of 11.5 and 17.7 days, a typical Kepler system. Compared to \texttt{TTVFast}, typical running time of \texttt{AnalyticLC} was roughly 5 times faster  when using the first or second order expansions. Gain in speed for the third order calculation was a factor of two, and the fourth order calculation did not yield a speed-up. The reason for this trend is that the number of disturbing function terms is not uniform among all orders. Therefore, in terms of efficiency, a significant gain relative to the state-of-the-art N-body integrator \texttt{TTVFast} is obtained when using the first or second order calculation.

\bibliography{main.bib}{}
\bibliographystyle{aasjournal}

\end{document}